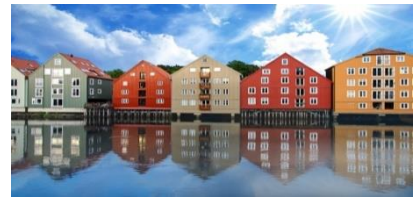

## Paper Information

| | |
|---|---|
| Paper number | 1391 |
| Paper title | An Advanced Cyber-Physical System Security Testbed for Substation Automation |
| Study Committee | SC D2 – Information systems, telecommunications and cybersecurity |
| Preferential subject | PS2: Technologies supporting the power grid for energy transition to carbon neutral energy production |
| Authors | HERATH Akila[1], LIU Chen-Ching[1], HONG Junho[2], GIRDHAR Mansi[2] |
| Affiliations (optional) | Virginia Polytechnic Institute and State University[1], University of Michigan Dearborn[2] |
| Country | USA |
| Email address | akilaasansana@vt.edu |

## Summary

In recent years, cyber attacks have increasingly targeted electric power grids, with substations being a primary focus. While the implementation of IEC 61850 in modern substations has improved power grid management efficiency, it has also exposed these systems to cyber vulnerabilities. The objective of cyber intrusions on substations is to manipulate communication between Intelligent Electronic Devices (IEDs) to operate circuit breakers (CBs), leading to disruptions in power grid operations. To strengthen substation security, considerable efforts are being directed toward developing strategies for cyber intrusion detection and mitigation. A Cyber-Physical System (CPS) testbed serves as a powerful platform for testing and validating these strategies.

This study introduces the design and development of a CPS testbed that incorporates widely used industry hardware from the substation's cyber layer, including IEDs, a remote terminal unit (RTU) and a time synchronization clock. Additionally, the testbed features an attacker module programmed with an open-source IEC 61850 library. This module can be used to develop realistic attack models on the IEDs in the testbed based on Sampled Value (SV) and Generic Object-Oriented Substation Event (GOOSE) messages. For instance, well known attack types include replay attacks, false data injections, and spoofing. Furthermore, a simulated IED that can replicate the functionalities of the real IEDs in the testbed is also included. This simulated IED is also programmed using an open-source IEC 61850 library. The physical layer is simulated using a Real-Time Digital Simulator which offers hardware-in-the-loop simulation capability. With the real-time modelling of the physical layer and the cyber layer devices enabling the measurement of information processing delays, real-time dynamics of the substation can be assessed. This level of observability of real-time dynamics is not available in many of the currently available testbeds with software-based simulation elements.

Cyber attacks exploiting SV and GOOSE protocols are demonstrated using the Hardware-in-the-Loop (HIL) testbed, along with an analysis of the Network Intrusion Detection System (NIDS), highlighting both its capabilities and limitations. Given that CB operation is a primary target of substation cyber attacks, realistic timing measurements are obtained from the testbed under SV and GOOSE-based scenarios. These simulations consider a range of potential device configurations, including evaluations with the simulated IED. Furthermore, an evaluation is conducted regarding the time frames in which detection and mitigation strategies should be deployed to block the malicious trip signals generated by an intrusion before reaching the CB.

Acknowledgement- This research is supported by the Office of Cybersecurity, Energy Security, and Emergency Response, Cybersecurity for Energy Delivery Systems Program, of the U.S. Department of Energy, under contract DE-CR0000021.

# 1  Introduction

The power system is undergoing a significant transformation by integrating Information and Communications Technology (ICT), enabling greater efficiency, reliability, and automation. As part of this transformation, substations are being modernized into digital substations that adopt the IEC 61850 standard for communication and control. This transformation offers numerous advantages, including interoperability between devices from different manufacturers, reduced wiring complexity with the use of Ethernet-based communication, and enhanced scalability for future upgrades. Digital substations improve operational efficiency by enabling real-time monitoring, fault detection, and automated responses to system disturbances.

While the IEC 61850 upgrade revolutionizes substations by enabling digital communication and automation, it has also introduced significant vulnerabilities to cyber attacks. The lack of built-in security measures for Ethernet-based protocols exposes substations to cyber threats. To address these challenges, substantial efforts are being made to develop robust detection and mitigation mechanisms that secure both the physical and cyber layers of substations. To enhance these efforts, Cyber-Physical System (CPS) testbeds for substations are being developed, enabling modelling and simulation of the interaction between physical devices and cyber systems under various attack scenarios.

In the literature, CPS testbeds with software-based implementations utilizing power systems and network simulation tools are commonly found [1]. Some testbeds incorporate real cyber-layer devices, enhancing realism [2]. Additionally, simulated or virtual Intelligent Electronic Devices (IEDs) have been utilized to evaluate cyber attack impacts [3]. However, these approaches may not accurately replicate the cyber-physical interactions in an actual substation environment. The CPS testbed presented in [4] aims to bridge this gap by closely mirroring a real substation environment. Many existing testbeds have the capability to simulate a wide range of cyber attacks. Nevertheless, it is critical for these testbeds to be enhanced to demonstrate their effectiveness in studying cyber attack detection and mitigation mechanisms involving real substation devices.

This paper presents a CPS security testbed for an IEC 61850-based substation, capable of real-time simulation with hardware-in-loop testing for both cyber and physical layers. The testbed can obtain realistic measures of the substation operation necessary for cyber security evaluation. In addition, the model of cyber attacks and detection of them within the testbed is demonstrated. In the remainder of the paper, section 2 discusses the security vulnerabilities of SV and GOOSE. Section 3 presents the testbed design whereas section 4 demonstrates the attacks. The experimental results are given in section 5 and section 6 concludes the paper.

# 2  Security Vulnerabilities in IEC61850-based SV and GOOSE

IEC 61850 is an international standard developed by IEC TC 57 for the automation of electrical substations, providing a comprehensive framework for communication, information modelling, and system architecture. The standard primarily defines three key communication protocols: Manufacturing Message Specification (MMS), Generic Object-Oriented Substation Event (GOOSE), and Sampled Values (SV). MMS operates over TCP/IP, functioning within the application, presentation, and session layers of the OSI model. It follows a client-server model and is primarily used for communication with station-level functions, such as supervisory control and monitoring. In contrast, GOOSE and SV are Layer 2 protocols that follow a publisher-subscriber communication model, making them suitable for time-critical functions within substations. Both protocols have strict time requirements and use multicast messages in the substation local area network (LAN). GOOSE is widely used for fast and reliable transmission of protection and control messages, while SV is essential for transmitting time-synchronized sampled measurement values.

Since GOOSE and SV are layer 2 protocols, they lack IP-level security mechanisms and do not have any built-in authentication or encryption in their standard forms (IEC 61850-8-1 for GOOSE and IEC 61850-9-2 for SV). If an adversary gains access to the substation LAN, the packets can be intercepted and injected back with modifications. Time-critical CB operations in the digital substation rely on SV and GOOSE communication as illustrated in Figure 1. Merging Unit IED (MUIED) receives current and voltage measurements from the current transformers (CTs) and voltage transformers (VTs) installed with the substation facilities.  The received measurements are converted into communication data packets and sent as SV to the protection and control IEDs (P&CIEDs). Subsequently, P&CIED processes the data, and if any discrepancy is detected, a GOOSE message is published to command a CB operation.  This message is received by MUIED, and it sends a trip signal to the CB. This CB operation can be manipulated by an adversary by injecting malicious SV or GOOSE messages into the substation LAN. However, the critical issue here is that such attacks can exploit vulnerabilities in communication protocols, potentially leading to incorrect breaker operations, system instability, or even cascading failures.

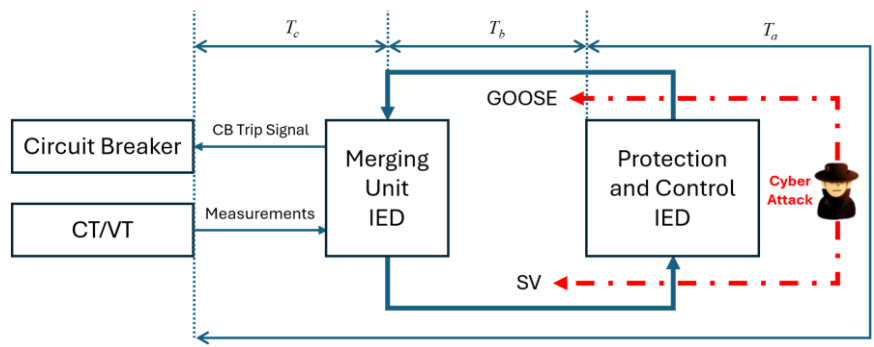

*Figure 1 – Illustration of CB operation with SV and GOOSE.*

Studies have been conducted on the detection of SV and GOOSE-based cyber attacks [5], [6]. Mitigation mechanisms proceed detection, and network switching based methods [7] have been proposed. However, research on real-time detection and mitigation of such attacks is still in its early stages. The total time $T_p$, needed to report fault data by SV and generate and transmit a GOOSE message to operate the CB, can be defined as in the equation below [2].

$$T_p = T_a + T_b + T_c \quad (1)$$

The time needed to receive measurements by MUIED, send them as SV, process the received data P&C IED, and send the GOOSE to the outgoing buffer for publishing is $T_a$, whereas $T_b$ is the time delay in the network communication for the GOOSE packet. $T_c$ is the time needed to save the GOOSE at the incoming buffer of the MUIED, process the data, send the CB trip signal, and operate the CB. If the deployed detection and mitigation mechanisms in the cyber layer are successful, a malicious trip signal will not be received by the physical CB. For the design and validation of such mechanisms, identification of the above-defined time frames is critical.

## 3 Design of the Testbed

### 3.1 Overview

The HIL testbed utilized in this project is built with industry-grade devices typical in an IEC 61850-based digital substation, as depicted in Figure 2. The physical layer of the power system is modeled using a real-time digital simulator. The merging unit IED, which is connected to the real-time digital via an amplifier, translates the hardwired communication signals into Ethernet-based data packets and vice versa. The data sent by the MUIED is received by two P&CIEDs. At the station level, a Remote Terminal Unit (RTU) with an integrated Human-Machine Interface (HMI) is deployed. The RTU serves as a communication node between the substation and the control center. A Global Positioning System (GPS) clock is used to synchronize substation devices with IEEE 1588 Precision Time Protocol (PTP). To simulate cyber threats, an attacker module is implemented. A network intrusion detection system (NIDS) [5] is also incorporated for detection testing. Additionally, a simulated IED, designed to replicate the functionality of the P&CIEDs, is included in the testbed. NIDS and simulated IED are programmed using an open-source IEC 61850 library on Linux computers.

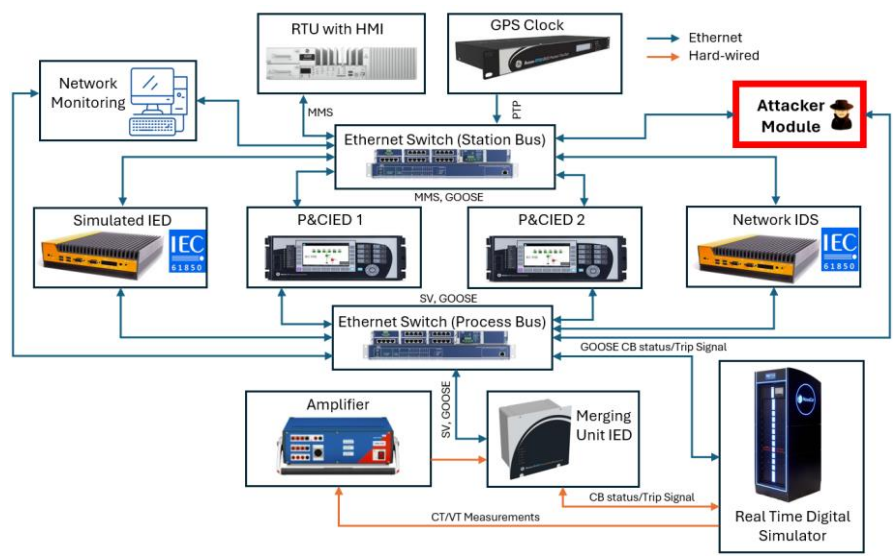

*Figure 2 – CPS security testbed for IEC 61850-based substation.*

### 3.2 Power System Model

The substation feeder setup is developed in developer software of the real time digital simulator as shown in Figure 3. This setup is used to study the cause-effect relationship of faults and cyber attacks interacting with real cyber layer devices. Scaled-down voltage and current signals from the feeder are sent out via the analog output interface of the simulator. The CB in the feeder (XCBR1) operates in two modes. It can operate as a conventional breaker which open/close based on the trip signal received from the MUIED. The digital interface is configured to receive the trip signal from the MUIED. Additionally, XCBR1 can mimic the operation of a breaker IED which operates with GOOSE received from P&CIED. The network interface of the simulator is configured to subscribe to the GOOSE trip signal of the P&CIED 1 and to publish the status of the CB as GOOSE messages.

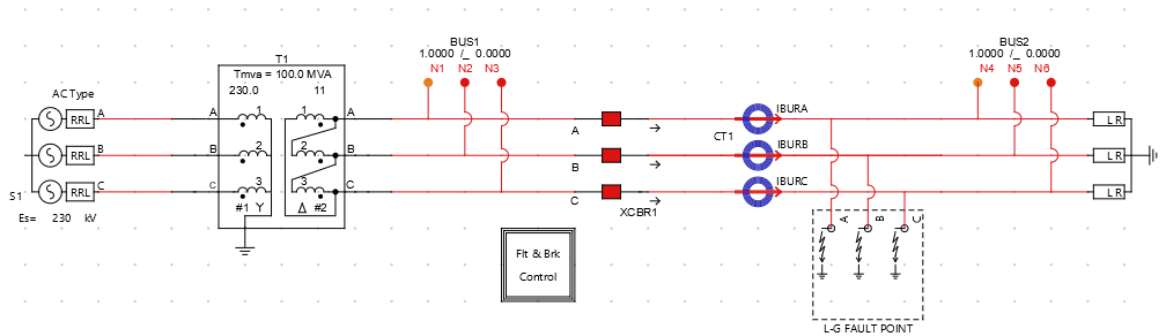

*Figure 3 – Power system model in the testbed.*

### 3.3    Simulated P&C IED

The P&CIEDs mainly cover the three aspects of protection, control, and measurements, with the inclusion of logical devices in each aspect. The Protection logical device houses the IEC 61850-based logical nodes such as PTOC (Time overcurrent protection) and PDIS (Distance protection). The logical nodes, including CSWI (Switch Controller), are part of the control logical device, whereas nodes such as MMXU (which measures voltage, current, and power) belong to the measurement logical device. The simulated P&CIED is developed to replicate these main functionalities of a P&CIED. The programming of the IED utilizes the SV subscriber and GOOSE subscriber/publisher application programming interfaces (APIs) defined in the library, with applying the necessary logic and calculations.

### 3.4    Network IDS

Rule-based algorithms based on the SV and GOOSE protocols are utilized in the network IDS in the testbed [5]. This unit is also programmed by utilizing the IEC 61850 library's SV subscriber and GOOSE subscriber/publisher APIs. The IDS is subscribed to all SV and GOOSE messages in the substation LAN. Each of the subscribed messages is monitored in separate threads, and rule-based algorithms are applied. When a rule violation is detected, a GOOSE message is sent out to report the intrusion.

## 4    Cyber Attack Simulation

### 4.1    SV Attack

The SV packet frame comprises several critical fields, including the destination address, source address, *ethertype*, Application Identifier (APPID), and packet length. The frame also contains the Application Protocol Data Unit (APDU), which encapsulates the SV data. The APDU consists of voltage and current measurements along with essential metadata fields such as *svID*, a system-wide unique identifier for the message, and *smpCnt*, a counter value representing the sample index within a transmission cycle. The semantics of the captured SV messages provide critical insights that may be leveraged to orchestrate future cyber attacks.

The instantaneous voltage or current value within an SV message, expressed as a function of *smpCnt*, can be modelled as the mathematical representation given in equation (2). Here, $X[n]$ denotes the sampled value at sample index *n* (corresponding to *smpCnt*), while $X_{peak}$ represents the peak amplitude of the signal. The sampling rate ($F_s$) is 4,800 samples per second for SV messages sending for protection calculations. Consequently, *smpCnt* ranges from 0 to 4799. The system frequency ($f$) and the phase angle ($\emptyset$) further define the waveform characteristics. In this testbed, SV false data injection (FDI) attacks are simulated by generating synthetic signal data with manipulated $X_{peak}$ values, capable of triggering undesired protection functions.

$$X[n] = X_{peak} \sin\left(2\pi \frac{f.n}{F_s} + \emptyset\right) \tag{2}$$

### 4.2 GOOSE Attacks

The frame structure of a GOOSE packet consists of fields similar to those found in an SV packet. The APDU includes the transmitted data that a potential attacker with access to the substation LAN, tries to manipulate. The *goID* field is a unique identifier of the message. In addition, *GoCBRef* and dataset fields also represent the GOOSE message type identity. The sequence numbers *stNum* and *sqNum* are changed with the change of the states represented by the GOOSE message and for each reception of a GOOSE message respectively. The timestamp of the message is updated with the change of *stNum*.

Replay attacks are conducted in the testbed by capturing GOOSE packets when the system is experiencing a fault and sending them back when the system is in normal operation. In addition, GOOSE spoofing is conducted by reinjecting modified GOOSE packets by inserting legitimate *stNum* and *sqNum* values. The control commands or the status information of the message can be altered.

## 5 Experimental Results

### 5.1 SV False Data Injection Attack Simulation

A malicious SV stream representing a peak current of 20kA was injected into the process bus following the semantics of the SV packets of the MUIED subscribed by P&CIED 1. The sampling interval of an SV packet when sending 4,800 data per second is 208 microseconds. However, it was required to send these packets at a sampling rate less than that to accommodate the processing delay at the attacker module. Synchronization between the original packets and the malicious packets of the SV stream with multicast MAC address Iec-Tc57_04:00:03 was achieved as shown in Figure 4(a). To assess the impact of this attack, the RMS currents calculated at P&CIED 1 were observed with its data logger interface illustrated in Figure 4(b). In the normal operation, the calculated RMS current is around 223 [A]. With the FDI attack, it raised to a value over 14 [kA], which is consistent with the injected false current data. This miscalculation at the P&C IED 1 led to trigger the overcurrent protection measures.

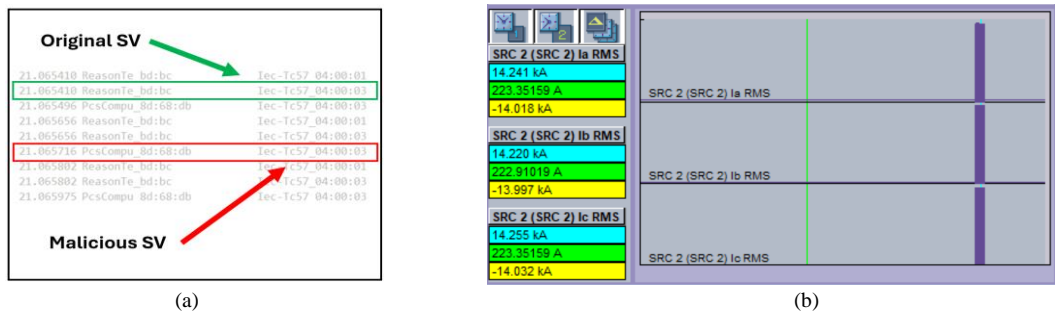

(a)  (b)

*Figure 4 – (a) Captured SV packets at the attack with receipt time, (b) Response shown in the data logger of P&CIED for the attack.*

### 5.2 GOOSE Replay Attack and Detection Simulation

A previously captured GOOSE packet with a 'True' boolean state to operate the CB was injected into the process bus. MUIED was tricked with the receipt of this malicious GOOSE packet and generated a CB trip signal while the system was not experiencing any fault conditions. Figure 5(a) illustrates this scenario in which the CB operated while the feeder current was normal. When the *stNum* and *sqNum* of the malicious packet and the last original GOOSE message are compared, clear mismatches can be observed. These mismatches are detected by the employed NIDS. NIDS detects the mismatch and publishes a GOOSE message to notify the intrusion. A measure of the NIDS detection time was obtained using a captured GOOSE packet, as shown in Figure 5(b). It was identified that the detection time is under 0.5 [ms]. However, this NIDS cannot detect cyber attacks that follow the protocol rules. A spoofing attack crafted by following the sequence numbers and sampling time frame rules cannot be detected.

### 5.3 Circuit Breaker Operation with IEC 61850

To determine the CB operation time using SV and GOOSE under various scenarios, line-to-ground faults were introduced at the fault point of the feeder model, as shown in Figure 3. With the MUIED operating alongside P&CIED 1, an average latency of approximately one cycle was recorded for CB tripping after fault occurrence, as illustrated in Figure 6(a). On the other hand, when the MUIED was used in conjunction with the simulated P&C IED, a slightly higher latency exceeding one cycle was observed, as shown in Figure 6(b). Specifically, the average operation times for the original IED and the simulated IED were 19 ms and 24 ms, respectively. Despite this 5 ms delay, the simulated IED effectively replicates near-real-time response behavior, making it a valuable tool for cybersecurity testing in HIL testbeds due to its open-source nature and enhanced operational visibility.

The two scenarios are compared: one in which the MUIED transmits the trip signal to the CB via a hardwired link, and another in which the CB functions as a breaker IED, bypassing the MUIED through a direct Ethernet link, both as shown in Figure 2. As illustrated in Figure 7(a), a delay between GOOSE message reception and CB operation was observed with MUIED, averaging 6 ms. However, when the CB directly operated as a breaker IED, this delay was reduced to 2 ms, as depicted in Figure 7(b). Consequently, the total average time from fault inception to CB operation decreased from 19 ms to 15 ms when the MUIED was bypassed, demonstrating improved response efficiency.

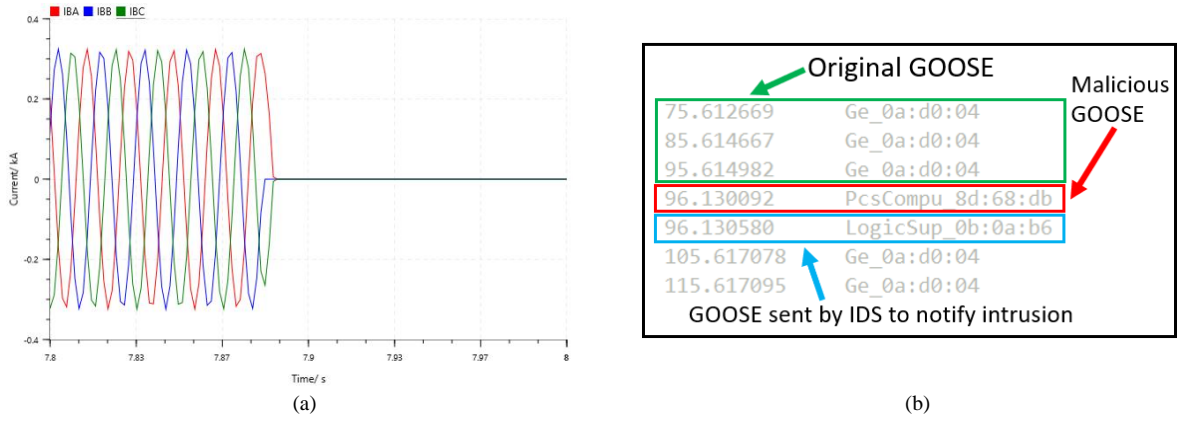

*Figure 5 – (a) CB operation by GOOSE replay attack, (b) Captured GOOSE at the attack with receipt time.*

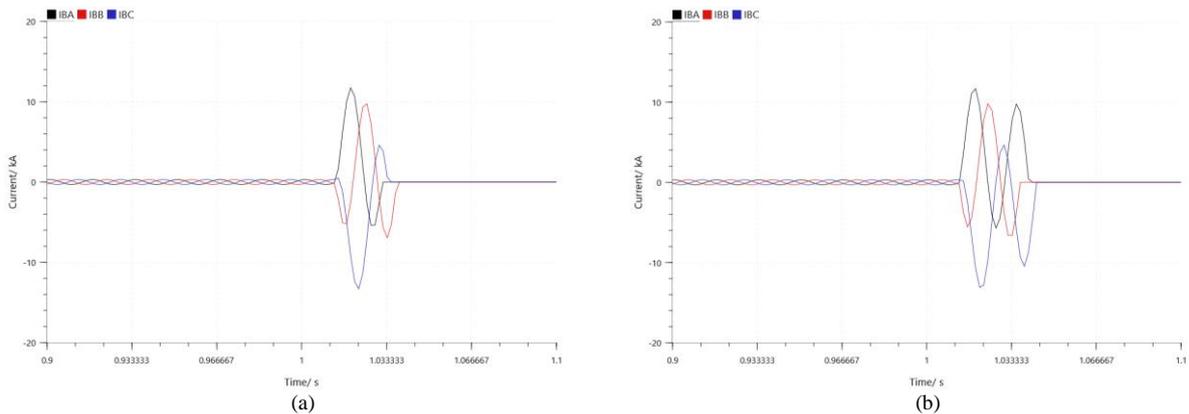

*Figure 6 – Feeder currents when CB is tripping with (a) P&CIED 1, (b) Simulated IED.*

### 5.4  Discussion on Detection and Mitigation Timeframes

The above observed results can be fitted into the parameters in (1) as, $T_a + T_b$= 13 ms, $T_c$= 6 ms. Although the maximum $T_b$ defined in IEC 61850 is 4 ms, the actual values can be in the microseconds range. In section 5.2, it is observed that the NIDS detection time plus the GOOSE transfer time was under 0.5 ms. Such that, $T_a \approx 13$ ms. For a successful FDI with SV, several packets may be required. The detection and mitigation for such an attack can afford time $T_a$, and it will be crucial to detect the attack within the first few malicious SV packets. Regarding GOOSE replay and spoofing attacks, time $T_c$ is available to detect the intrusion in the network and deploy the mitigation at the MUIED. Additional 4ms delay at the MUIED can be vital, as it extends the available timeframe compared to the operation with the breaker IED. The deployment of real-time detection and mitigation mechanisms within the identified narrow time ranges may be challenging with the currently available mechanisms. Further research is imperative to address these aspects comprehensively.

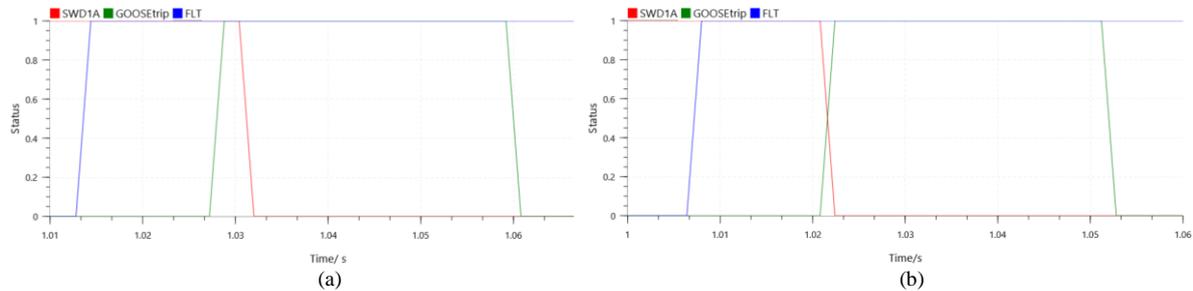

*Figure 7 – CB trip signal (SWD1A), GOOSE trip, and fault signal (FLT) for CB operation with (a) MUIED, (b) Breaker IED.*

## 6 Conclusions and Future Work

The proposed CPS security testbed, designed to closely emulate the operation of an IEC 61850-based substation, has demonstrated its capability to obtain realistic insights into cyber-physical interactions, which are crucial for cybersecurity study. A suite of SV- and GOOSE-based cyber attacks, executed through the attacker module of the testbed, was showcased alongside the operation of the NIDS integrated within the testbed. Time-critical measurements, essential for designing detection and mitigation strategies against SV and GOOSE attacks aimed at manipulating CB operations, were determined. Additionally, the response of a simulated IED which mimics the functions of a real P&CIED was demonstrated. Ongoing research focuses on leveraging the simulated IED to evaluate the functional-level impact of cyber attacks on IEDs. This is feasible due to the simulated IED's ability to replicate real functionalities while maintaining full observability. Furthermore, an advanced machine learning-based NIDS, trained on data generated by the testbed, will be integrated and evaluated. In parallel, a network-switching-based mitigation mechanism will be implemented. Additionally, the attacker module's capabilities will be expanded to more sophisticated attack scenarios.

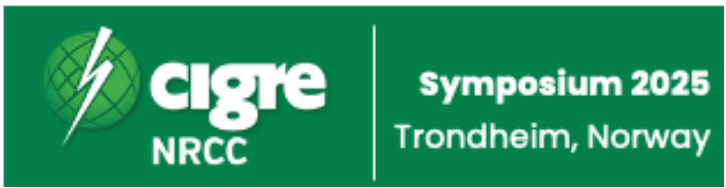 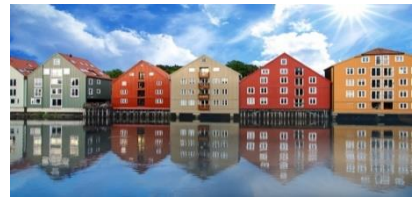